\documentclass[a4paper,12pt]{article}
\usepackage[colorlinks,citecolor=blue,linkcolor=blue,urlcolor=blue,filecolor=blue,backref=page]{hyperref}
\usepackage{authblk}
\usepackage{moreverb,url}
\usepackage{amsfonts}
\usepackage{authblk}
\usepackage{subfig}
\usepackage{graphicx}
\usepackage{longtable}
\usepackage{ragged2e}
\usepackage{natbib}
\usepackage[margin=1in]{geometry}
\providecommand{\keywords}[1]{{\small \textbf{\textit{Keywords---}}} #1}

\begin{document}
\title{Combining biomarker and self-reported dietary intake data:  a review of the state of the art and \\ an exposition of concepts.}

\author[1, 2]{Isobel Claire Gormley\thanks{claire.gormley@ucd.ie}}
\author[1, 3]{Yuxin Bai}
\author[3, 4, 5]{Lorraine Brennan}
\affil[1]{{\small School of Mathematics and Statistics, University College Dublin}}
\affil[2]{{\small Insight Centre for Data Analytics, University College Dublin}}
\affil[3]{{\small School of Agriculture and Food Science, University College Dublin}}
\affil[4]{{\small Institute of Food and Health, University College Dublin}}
\affil[5]{{\small Conway Institute, University College Dublin}}

\date{}
\maketitle
\begin{abstract}
	Classical approaches to assessing dietary intake are associated with measurement error.  In an effort to address  inherent measurement error in dietary self-reported data there is increased interest in the use of dietary biomarkers as objective measures of intake. Furthermore, there is a growing consensus of the need to combine dietary biomarker data with self-reported data. 

A review of state of the art techniques employed when combining biomarker and self-reported data is conducted. Two predominant methods, the calibration method and the method of triads, emerge as relevant techniques used when combining biomarker and self-reported data to account for measurement errors in dietary intake assessment. Both methods crucially assume measurement error independence. To expose and understand the performance of these methods in a range of realistic settings, their underpinning statistical concepts are unified and delineated, and thorough simulation studies conducted.

Results show that violation of the methods' assumptions negatively impacts resulting inference but that this impact is mitigated when the variation of the biomarker around the true intake is small. Thus there is much scope for the further development of biomarkers and models in tandem to achieve the ultimate goal of accurately assessing dietary intake. 

\end{abstract}


\maketitle

\section{Introduction}
A fundamental aspect of nutritional epidemiology is the ability to measure what people are eating. Over the years a number of well utilised self-reported tools or instruments have emerged to assess dietary intake. Commonly employed tools include food diaries, food frequency questionnaires (FFQ) and 24 hour recalls. The FFQ is possibly the most frequently employed tool; it ascertains a participant's usual dietary intake from a list of common foods over a defined period of time and is easy and inexpensive to administer to very large studies of participants. However the FFQ, as well as food diaries and 24 hour recall data, rely on participants accurately recalling food consumption over the specified time period and thus all have well documented associated measurement error \cite{bingham1995, carroll1998, heitmann1995, johansson1998, kirkpatrick2016, ocke2013, poppitt1998, pryer1997, thurigen2000}. Typical examples of measurement error include (but are not limited to) energy under-reporting, recall errors and difficulty in assessment of portion sizes \cite{bingham2002, kipnis2002}.  Such errors lead to reduced power, underestimated associations and false findings which may contribute to inconsistencies in the field of nutritional epidemiology \cite{dhurandhar2015, marshall1999}. 

In an effort to address such measurement error issues, there has been increased interest in the use of biomarkers in conjunction with the classical dietary data \cite{freedman2010}. Dietary biomarkers are found in biological samples and are related to ingestion of a specific food or food group \cite{gibbons2015, gibbons2015b}. Established biomarkers include sodium, urinary nitrogen, sucrose and doubly labelled water  for salt, protein, sugar and energy intake respectively \cite{kipnis2002, bingham1997, kipnis2003, tinker2011, tooze2004}. The use of such biomarkers to account for measurement error in self-reported data has been demonstrated in a number of studies \cite{schatzkin2003, subar2003, freedman2014, freedman2015}. Repeated measurements in conjunction with biomarkers have also been employed in an attempt to account for the inherent measurement error in self-reported data \cite{rosner2008}. Such repeated measurements are also suggested to be required when the nutrients under study are infrequently consumed and more complex measurement error models are required \cite{geelen2015, kipnis2009}. Approaches which combine multiple biomarkers for one nutrient with self-reported data have also been documented e.g. for total sugar intake \cite{tasevska2011}. 

Thus, in recent years there has been a substantial growth in the field of dietary biomarkers due to their potential to assist in accurately assessing dietary intake \cite{keogh2013, prentice2002}. With this in mind, the objective of this article is to explore and appraise the state of the art techniques that account for measurement error in dietary assessment by combining dietary biomarkers with self-reported data, while delineating the underlying statistical principles. While there is a wide, historic literature on measurement error \cite{carroll2006} most reviews thereof \cite{geelen2015, bennett2017, guolo2008, keogh2014} relate to the use of biomarkers and self-reported data to correctly infer diet-disease associations or in other such model extensions.  The scope herein differs due to its focus on the use of a biomarker based approach to account for measurement errors in the assessment of dietary intake. Further, a unifying and coherent exposition of the statistical detail underlying popular methods to combine biomarker and self-reported data is provided, with the aim of highlighting the fundamental method assumptions on which researchers rely and to highlight the sensitivity of such methods to assumption violations.

The article proceeds as follows: Section \ref{sec:notation} details a unifying notation while Section \ref{sec:literature} reports on the state of the art techniques used when combining self-reported dietary data and biomarker data as established by a review of the literature. Two techniques emerge as predominant: the calibration method and the method of triads and Section \ref{sec:stats} delineates the statistical details and assumptions often hidden behind their use. A comprehensive simulation study provides a clear exposition of the methods: illustrative examples of their performance in a range of realistic settings are provided and both techniques are subsequently critiqued. The R \cite{R18} code used to conduct the simulation study is available at \url{github.com/gormleyi}.  Finally, Section \ref{sec:discussion} concludes with an overview and recommendations.

\section{Notational framework}
\label{sec:notation}

Many studies in nutritional epidemiology take the form of large, population-based studies. Typically, in a large study with $N$ participants, participant $i$ may provide self-reported dietary intake data $W_i$ along with other covariate data $Z_i$. The dietary intake data is most commonly in the form of an FFQ, a 4 day food diary or 24 hour dietary recall data. Covariate data such as gender, ethnicity and age for example, are intuitively influential in dietary intake, and thus a range of such potentially important factors are recorded. Here the focus is on situations where such data are collected with the goal of assessing the true dietary intake for participant $i$, denoted $X_i$. In large studies $X_i$ is unobservable and thus the aim is to predict $\mathbb{E}(X_i |  W_i, Z_i)$ for $i = 1, \ldots, N$ i.e. to predict for each participant their dietary intake based on their self-reported and covariate data.

Interest has grown in the use of biomarkers in conjunction with self-reported data to account for measurement error in dietary assessment. Therefore, in a smaller sub-study with $n$ participants biomarker data $M_i$ for each participant $i$ will typically be recorded in addition to the self-reported and covariate data.  In some cases, such as feeding studies, the true dietary intake may be known for sub-study participants as controlled volumes of food are provided to participants. The expense of biomarker collection, and controlled feeding where applicable, means that typically  $n\ll N$.

Some instruments, known as reference instruments, are used in the collection of self-reported intake data and are deemed to be more accurate than others. While generic and the least onerous for participants to use, the FFQ is often deemed less accurate than the more detailed 4 day food diary or 24 hour recall data. The latter is often used as the reference instrument when estimating the accuracy of an FFQ, and thus herein is denoted $R_i$ for participant $i$.

In a concerted effort to account for dietary intake measurement error, many studies have complex experimental designs, for example collecting repeated measurements or employing sophisticated sampling strategies. Herein the focus is on a simple cross-sectional experimental design, to which the more complex studies can therefore relate. 

\section{Review of state of the art techniques}
\label{sec:literature}

A detailed exploration of the literature on state of the art techniques used when combining self-reported dietary intake with biomarker data was conducted through Google Scholar, pub-Med and Web of Science using the search terms `dietary measurement error biomarker', `biomarker calibration for dietary intake' and `dietary measurement error correction'. The resulting articles were explored as were references therein. When combining dietary biomarkers with self-reported data to account for measurement error in the assessment of dietary intake, the general consensus from the obtained articles is that the two most utilised techniques are (a) the calibration method and (b) the method of triads. An overview of the literature that avails of these methods, along with detail on the concepts underpinning them, is delineated in what follows. More generally, the literature includes studies which employ biomarker and self report data in a wide range of experimental settings and epidemiological considerations, utilising a variety of biomarkers; these studies often rely on the calibration method or the method of triads as an inherent part of their overall analysis. The interested reader is deferred to Tables \ref{tab:Calib} and \ref{tab:MoT}  in the Appendix which encapsulate some of the literature which employ biomarker and self report data for a variety of purposes.

\subsection{The calibration method}
\label{sec:calibration}

In the context of accounting for measurement error in dietary intake assessments, the calibration method refers to an approach that uses a biomarker to aid correction of the error in a less accurate self-reported dietary intake instrument. The calibration method is typically applied to data generated in a small sub-sample of a larger study, with the results then employed to account for measurement error in the self report data from the remainder of the study. It is important to note that the calibration method is distinct from `regression calibration' related methods which are extensively used approaches in the nutritional epidemiology literature to estimate associations between diet and other factors such as health outcomes \cite{rosner1989, spiegelman1997, frost2000, carroll2006, beydoun2007, freedman2008, freedman2011,   bennett2017}; some of these methods also rely on the use of biomarkers. The calibration method is often an initial, fundamental step in regression calibration where biomarkers are employed. As the focus herein is on correction of measurement error in self-reported data using biomarkers, the calibration method itself is explored.

Measurement errors can be classical, systematic, heteroskedastic or differential, and the error form may be related to an outcome of interest \cite{keogh2014}. As the aim here is to coherently explore methods applicable to a basic study to which other more complex (e.g. repeated measures) studies may relate, the focus is on a cross-sectional experimental design without a particular disease outcome or future model of specific interest. Thus throughout it is assumed that measurement errors follow the classical measurement error framework i.e. the truth is measured with additive error, usually with constant variance \citep{carroll2006}, as detailed in (\ref{eqn:classical}) below.

The calibration method assumes that the self-reported data $W_i$ (FFQ data, say) completed by participant $i$ in a sub-study is linearly related to their true dietary intake $X_i$ with additive random error i.e. 
\begin{eqnarray*}
W_i & = & \mu_W + \Lambda_{WX} X_i + \epsilon_{W_i}
\end{eqnarray*}
where $\mu_W$ is known as the \emph{constant scaling factor}. The \emph{proportional scaling factor} $\Lambda_{WX}$ is a measure of the strength (and direction) of the relationship between the FFQ and true intake. The FFQ specific error $\epsilon_{W_i}$ is assumed to be mean zero Normal error with variance $\sigma_W^2$. Further, the calibration method assumes that the biomarker measurement for sub-study participant $i$ is also linearly related to $X_i$ with additive random error. That is, for the biomarker the classical measurement error model holds where
\begin{eqnarray}
M_i & = & X_i + \epsilon_{M_i}                                                                        
\label{eqn:classical}
\end{eqnarray}
and $\epsilon_{M_i} \sim N(0, \sigma_M^2)$. In the sub-study $M_i$ and $W_i$ are recorded for $i = 1, \ldots, n$ but for the remaining study participants only $W_i$ is recorded. Thus the calibration method can be used to predict the dietary intake for the study participants conditional on their self-reported data; statistically the conditional distribution  $p(X_i | W_i)$ is required. Since the dietary intake is latent it is not possible to regress $X_i$ on $W_i$ to obtain this distribution. Instead $M_i$ is used as a surrogate for $X_i$ and is regressed on $W_i$ using the sub-study data where
\begin{eqnarray}
M_i & = & \beta_0 + \beta_1 W_i + \epsilon_i 
\label{eqn:meqn}
\end{eqnarray}
and $\epsilon_i \sim N(0, \sigma^2)$. The regression coefficients $\beta_0$ and $\beta_1$ are typically estimated using least squares. The predicted dietary intake for the study participant $i$ is then derived using the regression coefficient estimates:
\begin{eqnarray*}
\mathbb{E}(X_i | W_i) & = & \hat{\beta}_0 + \hat{\beta}_1 W_i 
\end{eqnarray*}
The calibration method works since $\mathbb{E}(X_i | W_i) = \mathbb{E}(M_i | W_i)$ under the crucial assumption that the errors $\epsilon_{W_i}$ and $\epsilon_{M_i}$ are independent (see Section \ref{sec:calibstats} for further details). Note that it is the estimate $\hat{\beta}_1$ that is often subsequently employed in regression calibration models to correctly account for attenuation in diet-disease associations \cite{keogh2014}.

It is well established that a participant's covariates may have an influence on their self-reported intake \cite{freedman2014, neuhouser2008, preis2011, prentice2013} and thus must be accounted for in the study design or in the resulting calibration. Typically, the calibration model is
\begin{eqnarray*}
W_i & = &  \mu_W + \Lambda_{WX} X_i + \gamma_{WZ} Z_i + \epsilon_{W_i}
\end{eqnarray*}
where $\gamma_{WZ}$ quantifies the relationship between $W_i$ and $Z_i$ conditional on the true dietary intake; the classical measurement error model for the biomarker (\ref{eqn:classical}) remains unchanged. The biomarker data are then regressed on the self-reported and covariate data with the resulting regression coefficients employed to derive $\mathbb{E}(X_i | W_i, Z_i)$.   

The calibration approach is prevalent throughout the literature. Typically researchers have included covariates such as BMI in the calibration method \cite{freedman2014, huang2013}, noting their inclusion is expected to strengthen the precision of the resulting calibration equations \cite{mossavar2013}. Prentice et al. \cite{prentice2011} use DLW and UN biomarkers to calibrate FFQ, 4 day food diary and 24 hour dietary recall data and explore the level of correlation between calibrated values for energy and protein intake and a set of key covariates, such as body mass index, age, and ethnicity. Further, they conclude that using the calibration method and any of these self assessment procedures, while accounting for such influential covariates, may yield suitable consumption estimates for epidemiology studies.  Later, the authors \cite{prentice2013} use DLW and the calibration method to assess the short-term total energy intake from self-reported data and biomarkers with covariates BMI, age and ethnicity again demonstrated as influential.  Mossavar-Rahmani et al. \cite{mossavar2015, mossavar2017} construct calibration equations using biomarkers to correct self-reported measures for energy, protein, sodium and potassium specifically in the Hispanic Community Health Study. Other work has expanded the calibration method to include person-specific bias such as in the case of sugar biomarkers \cite{spiegelman2005, tasevska2014}. While the calibration method has been demonstrated to work well for nutrients obtained from frequently consumed foods \cite{geelen2015} model extensions are required when working with nutrients obtained from episodically consumed foods \cite{kipnis2009, tooze2006, zhang2011}.

The statistical detail underlying the calibration method, and extensive simulation studies, are provided in Section \ref{sec:stats}, demonstrating the power and pitfalls of the calibration method in a range of realistic settings.

\subsection{The method of triads}
\label{sec:mot}

The validity of a set of self-reported data can be assessed through its \emph{validity coefficient} i.e. the correlation between the set of self-reported data (usually an FFQ data set) and the true intake. The magnitude of any loss of statistical power or bias in inference based on the set of self-reported data relates to this coefficient. The method of triads (MoT) is an approach to computing the validity coefficient which has its roots in factor and path analysis \cite{loehlin1998} and is closely related to structural equation models \cite{fraser2004, kaaks1997}. The MoT provides a popular, straight forward approach to computing the validity coefficient without recourse to estimating the specifics of the relationship between self-reported data and true intake. The method can be used to estimate the regression coefficient of true intake on self-reported intake, provided replicate biomarker data are available \cite{rosner2015}.

 The MoT requires three dietary measurements -- typically for participant $i$ these are an FFQ ($W_i$), a reference method such as 24 hour recall data ($R_i$) and a biomarker measure ($M_i$).  The method relies on two assumptions: (i) linearity between the three measurements and the true intake and (perhaps more fundamentally) (ii) independence between the three measurement errors. The benefit of incorporating biomarker data is that its errors are likely to be independent of those of the FFQ and the reference; possible sources of error in biomarker data are likely to be very different to those in self-reported data of habitual intake.
 
As true intake is unobserved, the MoT provides indirect estimation of the correlation between the true and self-reported intake by using the correlations between the three observed measurements. The validity coefficient ($VC_{WX}$) for the FFQ is
\begin{eqnarray}
VC_{WX} &  = & \sqrt{\frac{\rho_{MW} \rho_{WR}}{\rho_{MR}}} 
\label{eqn:vc}
\end{eqnarray}
where $\rho_{MW},  \rho_{WR}$ and $\rho_{MR}$ denote the three-pairwise correlations between the biomarker and FFQ, the FFQ and the reference and the biomarker and the reference respectively. 

The MoT is frequently used to assess the validity and robustness of an FFQ \cite{dixon2006, mcnaughton2005, mcnaughton2007}; Yokota et al. \cite{yokota2010} provide a wide review of its use in the literature. Kabagambe et al. \cite{kabagambe2001} use the method to assess the validity and reproducibility of an FFQ among Hispanic Americans, combining biomarkers with FFQ and 24-hour recall data. Fowke et al. \cite{fowke2002} employ the method of triads when assessing cruciferous vegetable consumption using multiple 24-hour recalls, a food-counting questionnaire and urinary dithiocarbamates excretion levels. In a similar vein Daures et al. \cite{daures2000} assess FFQ validity using a biomarker with reference to beta-carotene intake. While it seems intuitive that the biomarker errors are independent of the self-reported and reference errors, the same cannot be said with certainty for the errors between the self-reported data and the reference; in such cases the MoT is invalid \cite{geelen2015}. Fraser et al. \cite{fraser2005} approach this issue by considering two biomarkers to estimate FFQ validity. Rosner et al. \cite{rosner2008} allow for correlated errors between self-reported and reference data in a repeated-measures setting. 

Statistical detail underpinning the MoT is provided in Section \ref{sec:stats}, as are extensive simulation studies, demonstrating the merits and difficulties of the MoT in a range of realistic settings.

\section{Statistical concepts and simulation studies}
\label{sec:stats}

Prior to conducting thorough simulation studies to explore the calibration method and the MoT, the statistical concepts underpinning each method are presented. This statistical detail is fundamental to ensuring that the assumptions underlying the methods are well understood and thus that the methods are utilised correctly in practise. With this in mind a coherent and unifying statistical framework is developed. The performance of the methods across a range of realistic settings is then assessed through the simulation studies. For clarity, and as the focus is on a basic model to which more complex experimental designs can relate, the influence of covariates is assumed to have been accounted for and thus  covariates are omitted from the statistical development and simulation studies.

\subsection{The calibration method: statistical concepts}
\label{sec:calibstats}

The calibration method is employed where self-reported (typically FFQ) data have been recorded in a large study of $N$ participants, with both biomarker and self-reported data being obtained from a sub-study on $n$ participants, where $n \ll N$. In order to present the calibration method and highlight its assumptions, a generative factor analytic statistical framework is posited. For participant $i$ the true (log) dietary intake is unobserved and assumed to be $X_i \sim N(\mu_X, \sigma_X^2)$. The pair $(M_i,W_i)$ are assumed linearly related to $X_i$ through
\begin{eqnarray}
\left(\begin{array}{c}M_i \\ W_i \end{array}\right) & = & \left(\begin{array}{c} \mu_M \\ \mu_W \end{array}\right)+ \left(\begin{array}{c} \lambda_{MX} \\ \lambda_{WX} \end{array}\right) X_i  + \left(\begin{array}{c} \epsilon_{M_i} \\ \epsilon_{W_i} \end{array}\right) 
\label{eqn:calibeqn}
\end{eqnarray}
where $\mu = (\mu_M, \mu_W)$ denotes the biomarker and self-reported data constant scaling factors respectively. The proportional scaling factors $\Lambda = (\lambda_{MX}, \lambda_{WX})$  are measures of the strength and direction of the relationship between the biomarker data and the truth, and the self-reported data and the truth respectively. The biomarker and self-reported data errors are assumed to have a zero mean bivariate Normal distribution with covariance matrix $\Sigma$ where
\begin{eqnarray*}
\Sigma & = & \left(\begin{array}{lr}\sigma_M^2 & \sigma_M \rho_{MW} \sigma_W \\
\sigma_W \rho_{MW} \sigma_{M} & \sigma_W^2 \end{array} \right) 
\end{eqnarray*}
and $\rho_{MW}$ denotes the correlation between the errors in $M$ and the errors in $W$. Under these assumptions, integrating out the latent dietary intake gives the marginal distribution of $(M_i, W_i)$: a bivariate Normal distribution with mean $(\mu_M + \lambda_{MW} \mu_X, \mu_W +  \lambda_{WX} \mu_X)$ and covariance matrix $\Omega$ where
\begin{eqnarray}
\Omega & = & \left( \begin{array}{lr} \lambda_{MX}^2 \sigma_X^2 + \sigma_M^2 & \sigma_M \rho_{MW} \sigma_W + \lambda_{MX} \lambda_{WX} \sigma_X^2 \\
\sigma_W \rho_{MW} \sigma_M + \lambda_{MX} \lambda_{WX} \sigma_X^2 & \lambda_{WX}^2 \sigma_X^2 + \sigma_W^2\end{array} \right).
\label{eqn:omega}
\end{eqnarray}
As the object of interest is the predicted intake for each study participant given their FFQ, the distribution of interest is the conditional distribution $p(X_i |W_i)$. From (\ref{eqn:calibeqn}) it is clear that the conditional distribution $p(W_i | X_i) = N(\mu_W + \lambda_{WX} X_i, \sigma_W^2)$. Thus, invoking Bayes' Theorem and properties of the multivariate Normal distribution, so  is the distribution of interest
\begin{eqnarray*}
p(X_i | W_i) & = & N(\tilde{\beta}_0  + \tilde{\beta}_1 W_i, \tilde{\Psi})
\end{eqnarray*}
where $\tilde{\Psi} = \sigma_W^2 \sigma_X^2/(\sigma_W^2 + \lambda_{WX}^2 \sigma_X^2)$. In order to predict dietary intake from FFQ data, the key parameters of interest are $\tilde{\beta}_0$ and $\tilde{\beta}_1$. Their true values can be analytically derived as:
\begin{eqnarray}\nonumber
\tilde{\beta}_0 & = & (\sigma_W^2  \mu_X - \sigma_X^2 \lambda_{WX} \mu_W)/(\sigma_W^2 + \lambda_{WX}^2  \sigma_X^2)\\
\tilde{\beta}_1 & = & (\lambda_{WX} \sigma_X^2)/(\sigma_W^2 + \lambda_{WX}^2 \sigma_X^2)
\label{eqn:tildebeta}   
\end{eqnarray}
Clearly, in reality these true values cannot be computed as $X_i$ is latent. However, the conditional distribution of the manifest variables $p(M_i |W_i)$ can be analytically derived. Given (\ref{eqn:calibeqn}) and (\ref{eqn:omega}), their conditional distribution is $p(M_i | W_i ) = N(\beta_0 + \beta_1 W_i, \:\: \Omega_{11} -  \Omega_{12}^2 \Omega_{22}^{-1})$ where
\begin{eqnarray}\nonumber
\beta_0 & = & \frac{\sigma_W^2  (\mu_M + \lambda_{MX} \mu_X ) - \sigma_X^2  \lambda_{WX} (\lambda_{MX}  \mu_W -  \lambda_{WX} \mu_M) - \sigma_M \rho_{MW} \sigma_W (\mu_W + \lambda_{WX} \mu_X )}
{\sigma_W^2 + \lambda_{WX}^2 \sigma_X^2 }\\
\beta_1 & = & (\lambda_{MX} \lambda_{WX} \sigma_X^2 + \sigma_M \rho_{MW} \sigma_W)/(\sigma_W^2 + \lambda_{WX}^2 \sigma_X^2)      
\label{eqn:beta}                                
\end{eqnarray}
The estimates $\hat{\beta}_0$ and $\hat{\beta}_1$ are obtained by regressing $M_i$ against $W_i$, $i = 1, \ldots, n$ as in (\ref{eqn:meqn}).

Comparing (\ref{eqn:tildebeta}) to (\ref{eqn:beta}) it is clear that the parameters $(\tilde{\beta}_0, \tilde{\beta}_1)$ and $(\beta_0,\beta_1)$ are equal when the assumptions of the calibration method hold i.e. (i)  that the classical measurement error model (\ref{eqn:classical}) holds and so $\lambda_{MX} = 1$ and $\mu_M = 0$ and (ii) that the errors are uncorrelated and so $\rho_{MW} = 0$.  When these assumptions hold, then (\ref{eqn:tildebeta}) $=$ (\ref{eqn:beta}) and the regression estimates $(\hat{\beta}_0, \hat{\beta}_1)$ obtained from regressing $M_i$  on $W_i$ are also estimates of $(\tilde{\beta}_0, \tilde{\beta}_1)$, and thus can be used to predict dietary intake values from the FFQ data, $\mathbb{E}(X_i| W_i)$. 

\begin{figure}[h]
\begin{tabular}{cc}
\subfloat[$\rho_{MW} = 0$]{\label{fig:alpha0} \includegraphics[width=7.15cm, height=5.5cm]{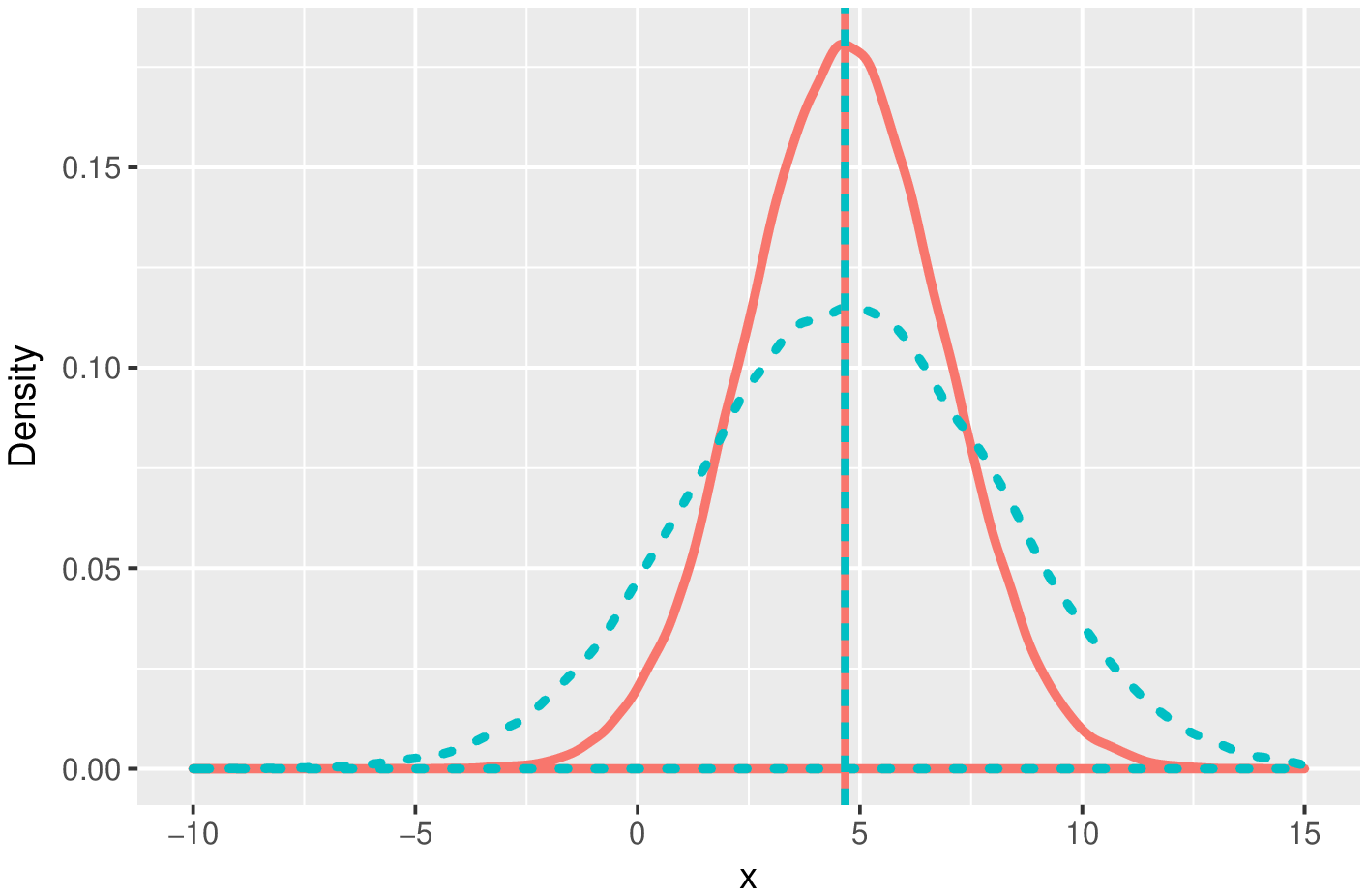}}&
\subfloat[$\rho_{MW} = 0.1$]{\label{fig:alpha0rho01} \includegraphics[width=7.15cm, height=5.5cm]{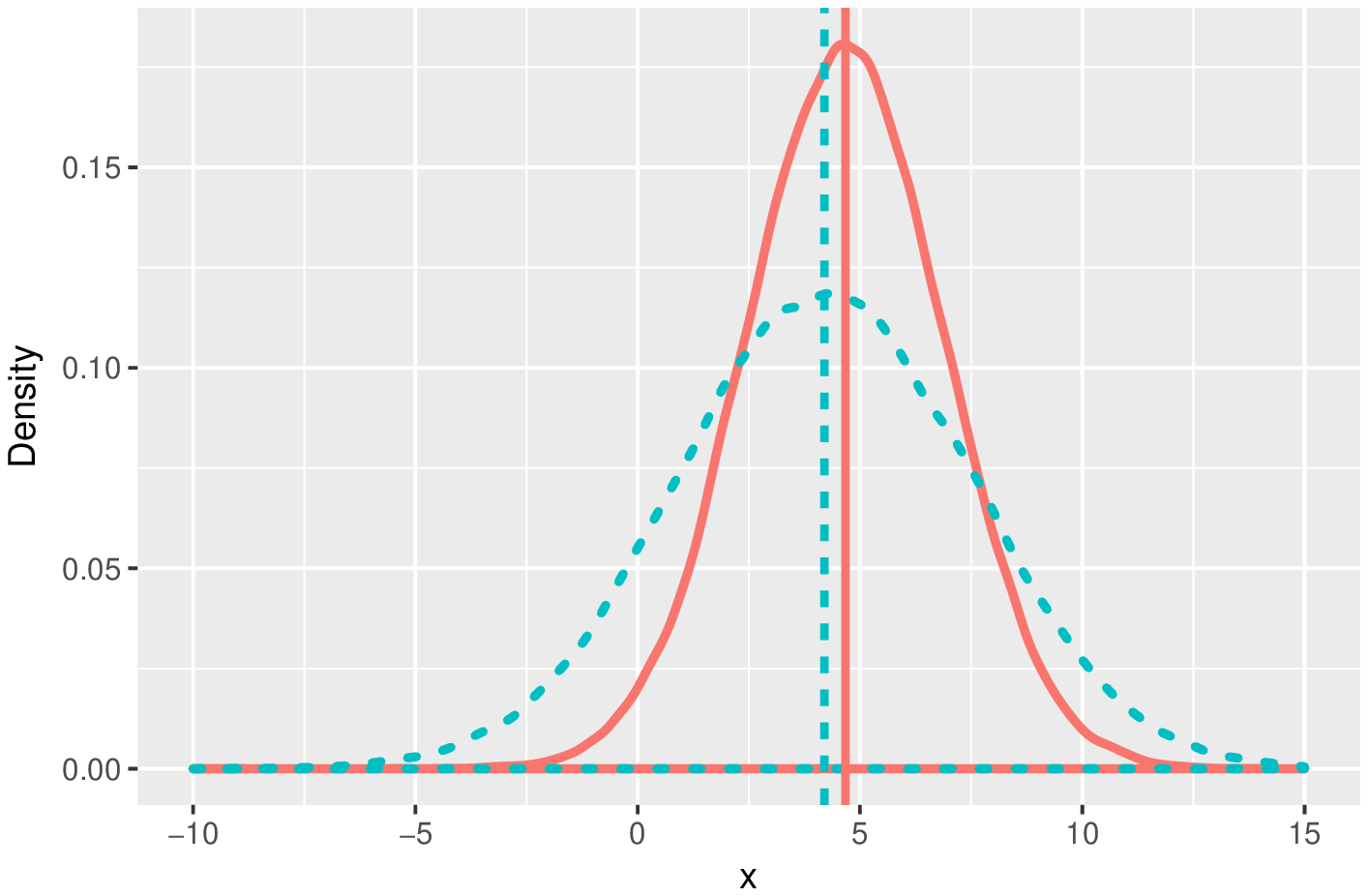}}\\
\end{tabular}
\caption{Kernel density estimates of simulated examples of $p(X_i |W_i)$ (solid red lines) and $p(M_i |W_i)$  (dashed blue lines) when (a) the calibration method assumptions hold and (b) when weak correlation is induced between the biomarker and FFQ errors. The vertical lines denote the distribution means. Here $\lambda_{WX} = 0.8$, representing that the FFQ and true intake are strongly related and $\sigma_X = \sigma_M = \log 14$. Further details on these settings are provided in Section \ref{sec:calibsim}.}
\label{fig:density}
\end{figure}

Figure \ref{fig:density} illustrates two simulated examples of $p(X_i |W_i)$ and $p(M_i |W_i)$; it is clear that the means of the distributions are the same when the calibration method assumptions hold, but when even weak correlation ($\rho_{MW} = 0.1$) is induced the means differ. Also visible is the difference in the variance of the distributions, even when the calibration method assumptions hold: the variance of $p(M_i | W_i)$ can be shown to be $\Omega_{11} - \Omega_{12}^2 \Omega_{22}^{-1} =  \tilde{\Psi} + \frac{\lambda_{WX}^2 \sigma_X^2 \sigma_M^2 + \sigma_W^2 \sigma_M^2}{\sigma_W^2 + \lambda_{WX}^2 \sigma_X^2}$ i.e. the variance $\tilde{\Psi}$ of $p(X_i | W_i)$ plus a positive term, and hence the increased variation in $p(M_i|W_i)$ compared to $p(X_i | M_i)$ observed in Figure \ref{fig:density}.

It should be additionally noted however that variance of the distribution of the \emph{predicted} intake $\mathbb{E}(X_i | W_i)$ suffers from the `shrinkage effect' well explored in the nutritional epidemiological literature \cite{kaaks1995, ferrari2004}. That is, given $\mathbb{E}(X_i | W_i) = \tilde{\beta}_0 + \tilde{\beta}_1 W_i$ and assuming the measurement error model (\ref{eqn:classical}) holds and  that the errors of $W$ and $M$ are independent, then
\begin{eqnarray*}
Var[\mathbb{E}(X_i | W_i )] & = & \tilde{\beta}_1^2 Var(W_i) =  \frac{\sigma_X^2}{1 + \sigma_W^2/\lambda_{WX}^2\sigma_X^2}
\end{eqnarray*}
Thus the larger the error variability $\sigma_W^2$ and/or the lower the proportional scaling factor $\lambda_{WX}$ the greater the shrinkage of variance of the distribution of the predicted versus true intakes.

\subsection{The calibration method: simulation study}
\label{sec:calibsim}

A simulation study is conducted to explore the performance of the calibration method across a range of settings. The statistical framework developed in Section \ref{sec:calibstats} is used to generate data from large studies with $N = 10,000$ and their related sub-studies with $n = 1000$. In order to emulate realistic nutrition studies, the parameter settings are extracted from Prentice et al. \cite{prentice2011} which reports the geometric mean for uncalibrated energy intake as estimated by an FFQ, and for calibrated energy intake using nutritional biomarker data. Thus, the true (log) intakes are simulated as $X_i \sim N(\mu_X = \log 2141, \sigma_X^2 = (\log14)^2)$.
 
The biomarker data adhere to the classical measurement error model (\ref{eqn:classical}) so $\mu_M = 0$ and $\lambda_{MX} = 1$. The parameter $\sigma_M^2$ which controls the variation of the biomarker measurement around the true intake is here considered to be related to $\sigma_X^2$ where $\sigma_M = \alpha \sigma_X$. Three different settings for $\alpha$ are considered $\alpha = \{0.5, 1, 2\}$. Thus the `best case scenario' is $\alpha = 0.5$, in which the biomarker does not vary much around the true intake. In contrast, the `worst case scenario' is $\alpha = 2$ where the biomarker is much more variable than the truth.

The self-reported data are considered with recourse to Prentice et al.  \cite{prentice2011}, with parameter settings $\mu_W=  \log1485$ and $\sigma_W =  \log 28$. The proportional scaling factor $\lambda_{WX}$ is a measure of the strength and direction of the relationship between the FFQ and true intake which in reality may vary depending on the nutrient or hypothesis under study. Thus, three settings for $\lambda_{WX}$ are considered: $\lambda_{WX} = \{0.1, 0.5, 0.8\}$ where the `best case scenario' is $\lambda_{WX} = 0.8$ in which there is a strong positive relationship between the FFQ and the truth. In contrast, the `worst case scenario' is $\lambda_{WX} = 0.1$ where there is little link. Finally, the correlation $\rho_{MW}$ between the errors in $M_i$ and $W_i$ requires consideration. Again, three settings are considered relating to the calibration method requirement of $\rho_{MW} = 0$ and a weak ($\rho_{MW} = 0.1$) and strong ($\rho_{MW} = 0.8$) violation of this assumption.

For each of the nine combinations of $\alpha$ and $\lambda_{WX}$ and for each setting of $\rho_{MW}$, a total of 500 sub-study data sets (consisting of $(X_i, W_i, M_i)$  for $i = 1,\ldots, n$) and 500 associated large study data sets (consisting of $(X_i, W_i)$ for $i = 1,\ldots, N$) are simulated. For each sub-study data set, the biomarker data are regressed on the self-reported data to produce estimates $(\hat{\beta}_0, \hat{\beta}_1)$; these estimates are used to derive $\mathbb{E}({X}_i |W_i)$ for each participant $i$. As the data are simulated, the true parameter values $(\tilde{\beta}_0, \tilde{\beta}_1)$ can be analytically derived and the true intakes for the $N$ study participants are known, facilitating assessment of the performance of the calibration method across the range of settings.

\begin{figure}
\centering
\subfloat[$\rho_{MW} = 0$]{\label{fig:betaratioA} \includegraphics[width=15cm, height=5cm]{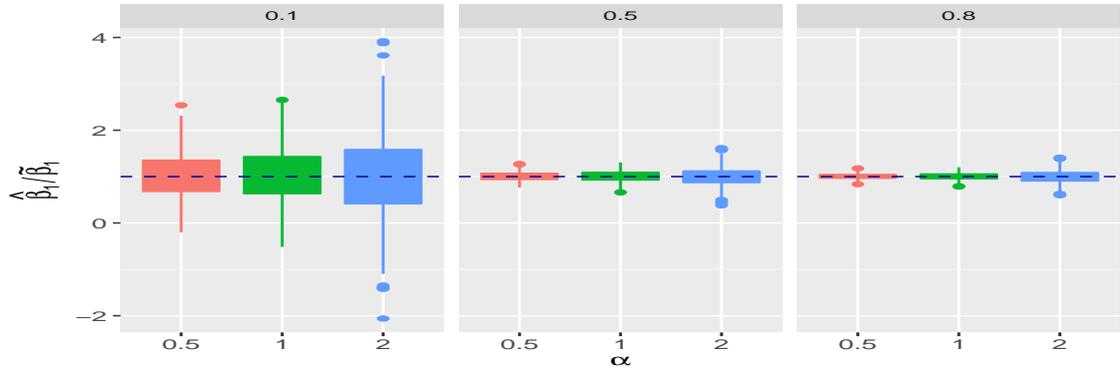}}\\
\subfloat[$\rho_{MW} = 0.1$]{\label{fig:betaratioB} \includegraphics[width=15cm, height=5cm]{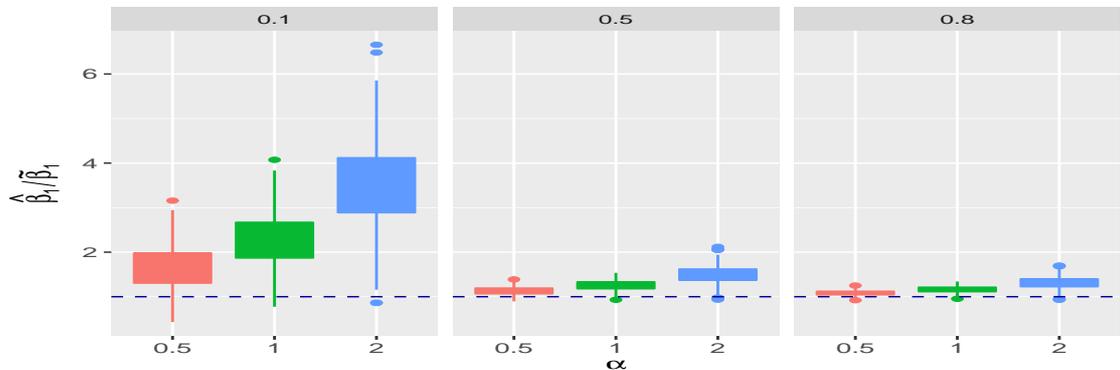}}\\
\subfloat[$\rho_{MW} = 0.8$]{\label{fig:betaratioC} \includegraphics[width=15cm, height=5cm]{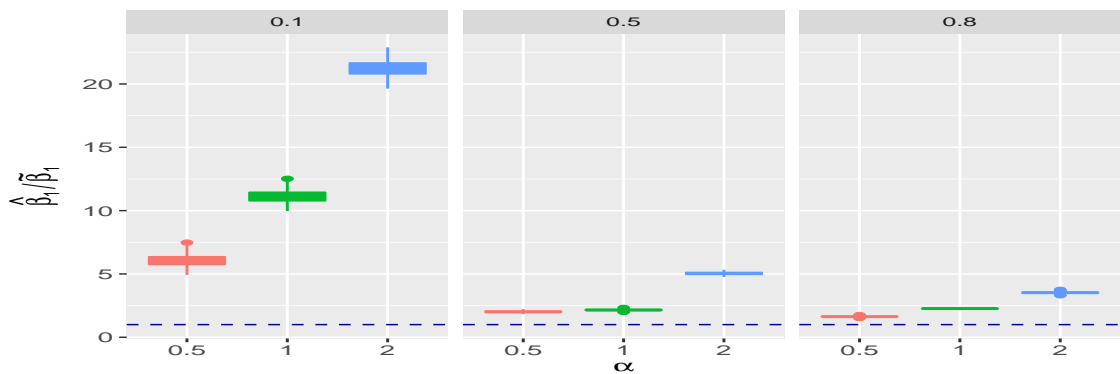}}
\caption{Estimates of the ratio $\hat{\beta}_1/\tilde{\beta}_1$ across 500 simulated sub-study datasets with (a) independent errors  (b) weakly correlated errors and (c) strongly correlated errors. The value of $\lambda_{WX}$ is denoted at the top of each subfigure, and the dashed horizontal line marks the line of equality.}
\label{fig:betaratio}
\end{figure}

Figure \ref{fig:betaratio} illustrates estimates of the ratio $\hat{\beta}_1/\tilde{\beta}_1$  across 500 simulated sub-study datasets for different values of $\alpha$ and $\lambda_{WX}$ and across the different settings of $\rho_{MW}$. When the assumptions of the calibration method hold (Figure \ref{fig:betaratioA}) the calibration method does very well across the range of settings. The ratio is close to one in general and while there is increased variation when the biomarker has large variance around the true value ($\alpha = 2$) and when the FFQ and the truth are weakly related ($\lambda_{WX} = 0.1$), the calibration method does consistently well. 

Figure \ref{fig:betaratioB} however shows the estimated ratio when weak correlation $(\rho_{MW} = 0.1)$ is present between the errors. As expected, given the statistical derivation above, the estimate $\hat{\beta}_1$ is no longer also an estimate of the true value $\tilde{\beta}_1$ and their ratio consistently deviates from one. The ratio is almost always greater than 1 (note the change in the vertical axis compared to Figure \ref{fig:betaratioA}), due to the increase in the numerator of $\beta_1$ in (\ref{eqn:beta}) compared to (\ref{eqn:tildebeta}). Also, there are now larger discrepancies when the biomarker is more variable than the truth (e.g. $\alpha = 2$) than was evidenced in Figure \ref{fig:betaratioA}. Given that the errors are only weakly correlated here, the scale of discrepancy is notable: in the best case scenario $\lambda_{WX} = 0.8, \alpha = 0.5$ in Figure \ref{fig:betaratioB} the largest ratio $\hat{\beta}_1/\tilde{\beta}_1 = 1.25$. In the realistic `worst case scenario' where $\lambda_{WX} = 0.1, \alpha = 2$ the estimate of $\beta_1$ is 6.66 times the true value in the most extreme case. This behaviour is exacerbated when the degree of correlation is increased to $\rho_{MW} = 0.8$. The discrepancy behaviour in Figure \ref{fig:betaratioC} is much more extreme than previously: for the best case $\lambda_{WX} = 0.8, \alpha = 0.5$ the maximum ratio is 1.76, while for the worst case $\lambda _{WX} = 0.1, \alpha = 2$ the estimate is 22.89 times the true value in the most extreme case. The propagation of such regression coefficient estimates into future analyses would clearly result in dubious inference.

\begin{figure}
\centering
\subfloat[$\rho_{MW} = 0$]{\label{fig:MAEa} \includegraphics[width=15cm, height=5cm]{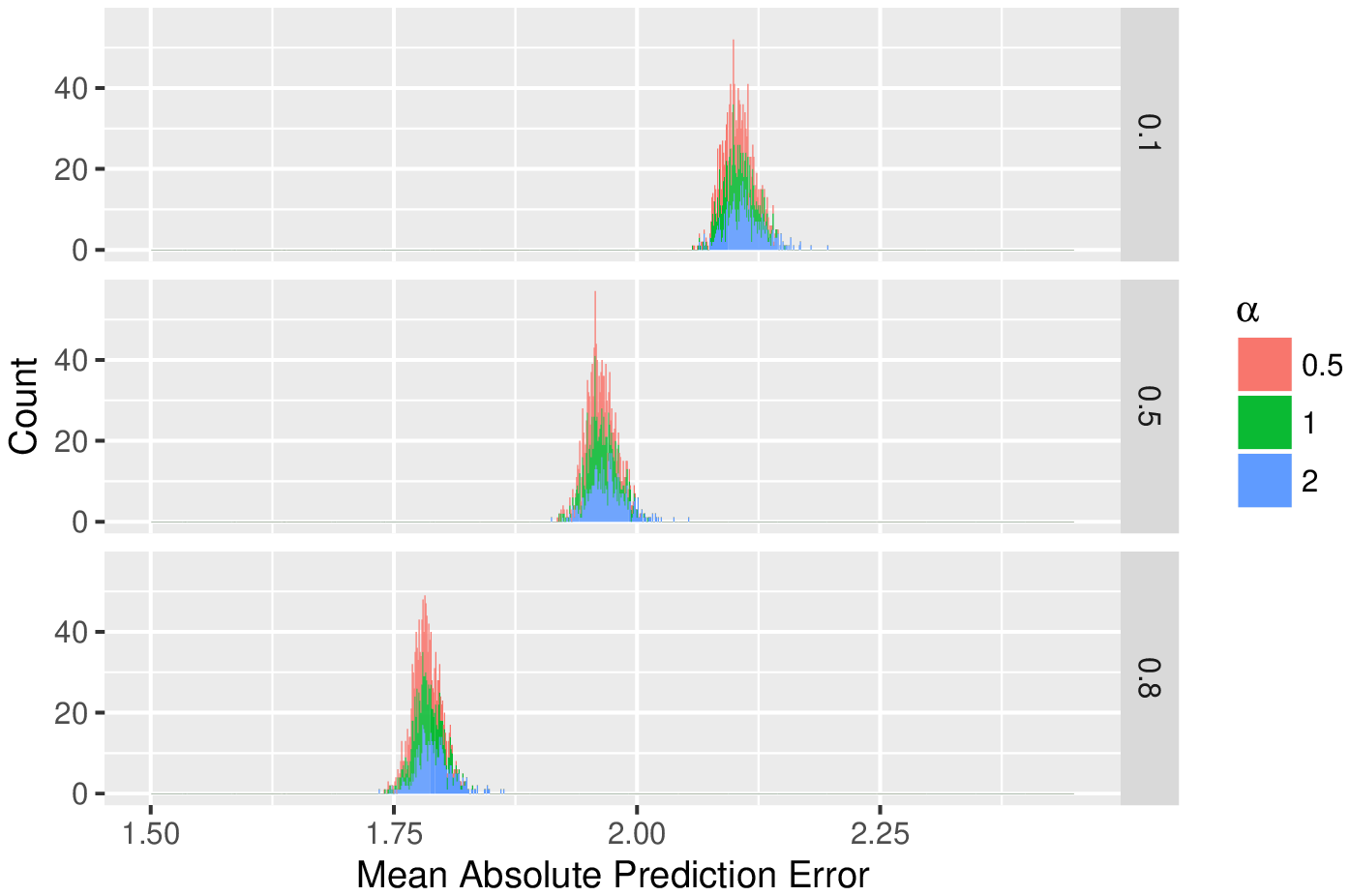}}\\
\subfloat[$\rho_{MW} = 0.1$]{\label{fig:MAEb} \includegraphics[width=15cm, height=5cm]{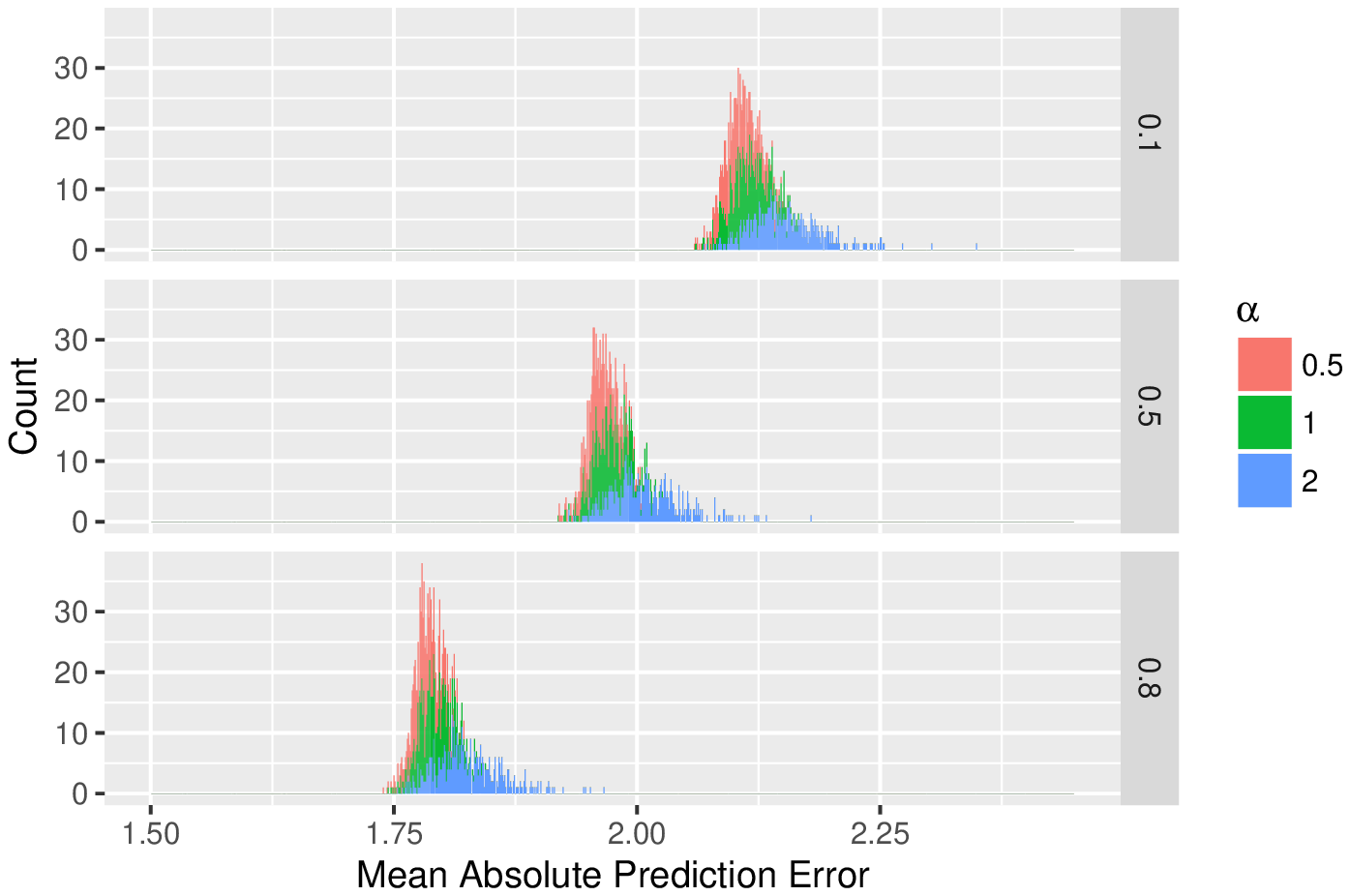}}\\
\subfloat[$\rho_{MW} = 0.8$]{\label{fig:MAEc} \includegraphics[width=15cm, height=5cm]{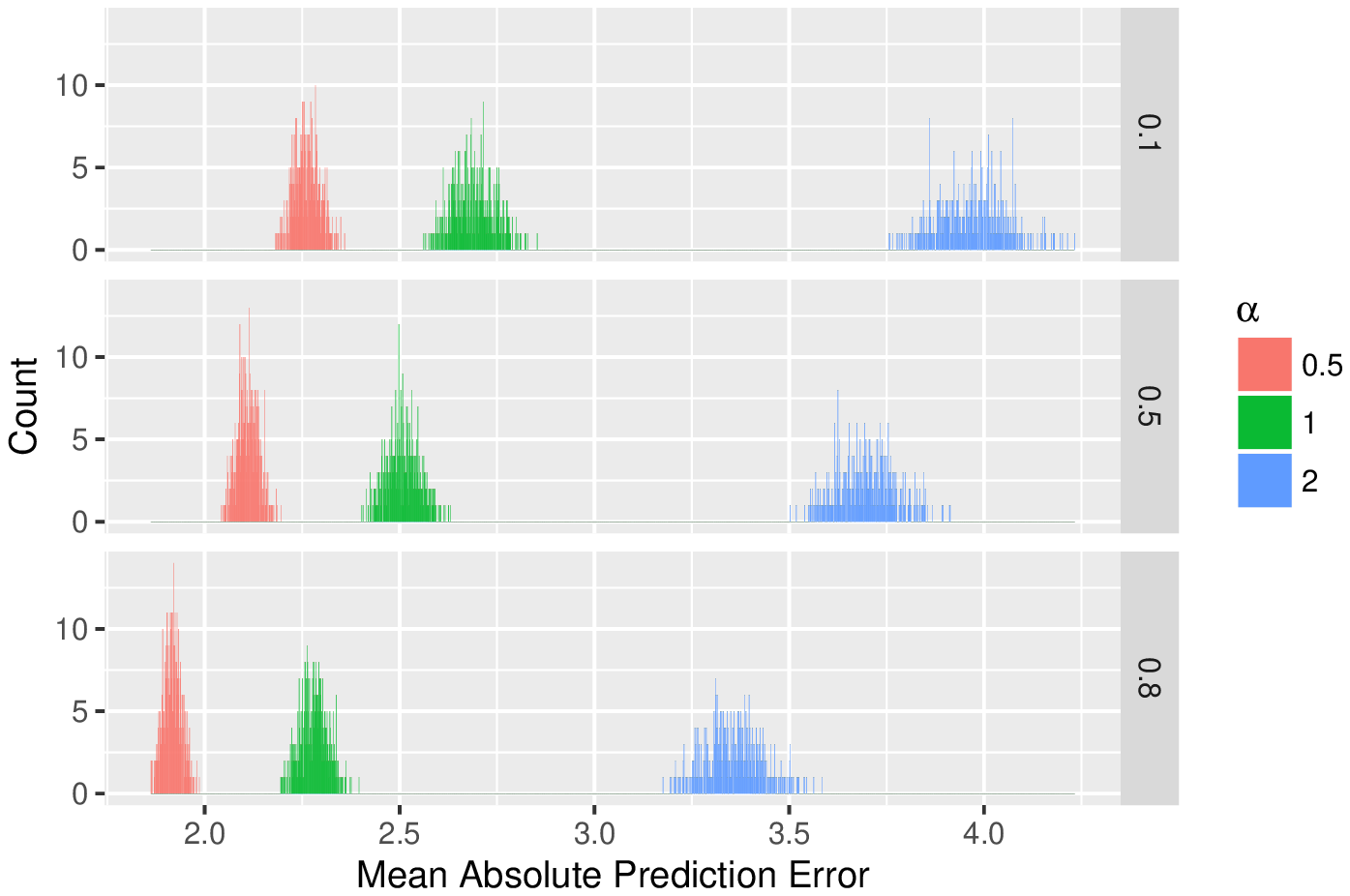}}
\caption{Mean absolute prediction errors across 500 simulated study datasets with (a) independent errors  (b) weakly correlated errors  and (c) strongly correlated errors. The value of $\lambda_{WX}$ is denoted on the right of each figure. The horizontal axes of (\ref{fig:MAEa}) and (\ref{fig:MAEb}) have been aligned for comparative purposes, but not that of (\ref{fig:MAEc}) for clarity.}
\label{fig:MAE}
\end{figure}

While estimating the regression coefficients is at times important for addressing attenuation bias in future disease-diet analyses, in some cases the predictions of dietary intake from the FFQ data are the entities of interest. Figure \ref{fig:MAE} demonstrates the mean absolute error between the intakes predicted from the FFQ data and the true known intakes across 500 simulated study data sets for $\rho_{MW} = \{0, 0.1, 0.8\}$. At first glance, the calibration method does not perform as impressively in terms of prediction: the mean absolute prediction errors are far from zero even in the `best case scenario' of $\alpha =0.5, \lambda_{WX} = 0.8$ and where the assumption of non-correlated errors holds (Figure \ref{fig:MAEa}). However, the prediction performance does not vary hugely when the independent errors assumption is weakly violated, as evidenced by the comparison of Figures \ref{fig:MAEa} and \ref{fig:MAEb}, for which the horizontal axes have been aligned for comparative purposes. While the variation in the prediction errors increases in the $\rho_{MW} = 0.1$ setting, the bias remains similar to that observed in the $\rho_{MW} = 0$ setting. Further, the variation in prediction errors is greater at higher levels of $\alpha$ than at lower levels of $\alpha$, which was not manifested in the uncorrelated errors setting. Additionally in Figure \ref{fig:MAEc} where the error correlations are strong, the prediction errors can be considerable, yet in the case of $\alpha = 0.5,\rho_{WX} = 0.8$ the prediction errors are in line with those achieved in the non-correlated error setting. 

Synoptically, if correlated errors are present, predictions can in general be of similar quality to those produced in the case of non-correlated errors, conditional on the biomarker having low variance around the true dietary intake, and somewhat independently of the strength of the relationship between the FFQ and the truth. Thus, when interest lies in predicting intake, a biomarker that does not vary much more than the truth and FFQ data that are strongly related to the truth, can mitigate the negative impact of the violation of the calibration method error assumptions. 

One final note is the crucial, underpinning classical measurement error assumptions $\mu_M = 0$ and $\lambda_{MX} = 1$ which fundamentally ensure the accuracy of the calibration method, even when measurement errors are independent. Any variation in these settings will notably impact the method's performance as equality of $(\tilde{\beta}_0, \tilde{\beta}_1)$ in (\ref{eqn:tildebeta}) and $(\beta_0,\beta_1)$ in (\ref{eqn:beta}) no longer holds.

\subsection{The method of triads: statistical concepts}
\label{sec:MoTstats}

When both self report data and dietary biomarkers are available, the method of triads is frequently employed to assess the validity of the set of self-reported data. The MoT requires three measurements to be made on each sub-study participant, usually the self-reported FFQ data $W_i$, reference 24 hour recall data $R_i$ and biomarker data $M_i$. The true (log) dietary intake is unobserved and is modelled as $X_i \sim N(\mu_X, \sigma_X^2)$. Similar to (\ref{eqn:calibeqn}), the MoT assumes the observed measurements are linearly related to the latent intake:

\begin{eqnarray}
\left(\begin{array}{c}M_i \\ R_i \\ W_i \end{array}\right) & = & \left(\begin{array}{c} \mu_M \\ \mu_R \\ \mu_W \end{array}\right) + \left(\begin{array}{c} \lambda_{MX} \\ \lambda_{RX} \\ \lambda_{WX} \end{array}\right) X_i  + \left(\begin{array}{c} \epsilon_{M_i} \\ \epsilon_{R_i} \\ \epsilon_{W_i} \end{array}\right) 
\label{eqn:MoTeqn}
\end{eqnarray}
where $\mu = (\mu_M, \mu_R, \mu_W)$ are the biomarker, reference and self-reported constant scaling factors respectively. The proportional scaling factors $\Lambda = (\lambda_{MX}, \lambda_{RX}, \lambda_{WX})$ are measures of the strength (and direction) of the relationship between the three observed measurements and the true latent intake. The trivariate errors $\epsilon = (\epsilon_{M_i}, \epsilon_{R_i}, \epsilon_{W_i})$ follow a mean zero Normal distribution with covariance $\Sigma$:
\begin{eqnarray}
\Sigma & = & \left(\begin{array}{lcr}
\sigma_M^2 & \sigma_M \rho_{MR} \sigma_R & \sigma_M \rho_{MW} \sigma_{W}\\
\cdot & \sigma_R^2 & \sigma_R \rho_{RW} \sigma_{W} \\
\cdot & \cdot & \sigma_{W}^2\end{array} \right) 
\label{eqn:SigmaMoT}
\end{eqnarray}
Respectively, $\rho_{MR}$ and $\rho_{RW}$ denote the correlation between the errors of the biomarker and the reference and between the reference and the self-reported data. Marginally, the covariance matrix of $(M_i, R_i, W_i)$ is $\Omega$ where
\begin{eqnarray}
\Omega & = & \left( \begin{array}{lcr} 
\lambda_{MX}^2 \sigma_X^2 + \sigma_M^2 & \sigma_M \rho_{MR} \sigma_R + \lambda_{MX} \lambda_{RX} \sigma_X^2 & \sigma_M \rho_{MW} \sigma_W + \lambda_{MX} \lambda_{WX} \sigma_X^2 \\
\cdot & \lambda_{RX}^2 \sigma_X^2 + \sigma_R^2 & \sigma_R \rho_{RW} \sigma_W + \lambda_{RX} \lambda_{WX} \sigma_X^2\\
\cdot & \cdot & \lambda_{WX}^2 \sigma_X^2 + \sigma_W^2\end{array} \right)
\label{eqn:omegaMoT}
\end{eqnarray}

The MoT involves computing $VC_{WX}$ the validity coefficient of the FFQ and the true intake:
\begin{eqnarray*}
VC_{WX} & = & Cor(W, X) = Cov(W, X)/( \sqrt{Var(W)} \sqrt{Var(X)})
\end{eqnarray*}
In the simulation setting the true value $\tilde{VC}_{WX}$ can be analytically derived. It can be shown \cite{kaaks1997}, given the properties of the factor analytic model (\ref{eqn:MoTeqn}), that $Cov(W, X) = \lambda_{WX} \sigma_X^2$ and extracting $Var(W)$ from $\Omega$ in (\ref{eqn:omegaMoT}) gives:
\begin{eqnarray}
\tilde{VC}_{WX} & = & \lambda_{WX} \sigma_X^2 / (\sqrt{\lambda_{WX}^2 \sigma_X^2 + \sigma_W^2} \sqrt{\sigma_X^2})
\label{eqn:VCtrue}
\end{eqnarray}
In reality $X_i$ is latent and the MoT is used to estimate $VC_{WX}$. The method crucially assumes that the correlation between the errors in each measure is 0 i.e. $\rho_{MR} = \rho_{MW} = \rho_{RW} = 0$. In this case, from $\Omega$ in (\ref{eqn:omegaMoT}), the covariance between measures $j, k = \{M_i, R_i, W_i\}$ simplifies to $Cov(j, k) = \lambda_{jX} \lambda_{kX} \sigma_X^2$. The correlation between each pair of measures is then $\rho_{jk} = \lambda_{jX} \lambda_{kX} \sigma_X^2/(\sqrt{Var(j)} \sqrt{Var(k)})$ and it follows that
\begin{eqnarray*}
\frac{\rho_{MW} \rho_{WR}}{\rho_{MR}} & = & 
\frac{\lambda_{MX} \lambda_{WX} \sigma_X^2}{\sqrt{Var(M)} \sqrt{Var(W)}} \frac{\lambda_{WX} \lambda_{RX} \sigma_X^2}{\sqrt{Var(W)} \sqrt{Var(R)}}  \frac{\sqrt{Var(M)} \sqrt{Var(R)}}{\lambda_{MX} \lambda_{RX} \sigma_X^2} \\
 & =  & \frac{\lambda_{WX}^2 \sigma_X^2}{Var(W)}
 \end{eqnarray*}
the square root of which is the true value of the validity coefficient (\ref{eqn:VCtrue}). (Note that if any of the correlations are negative, or the estimated correlation is greater than 1 (known as a Heywood case), the MoT cannot numerically proceed \cite{ocke1997}.) Thus, the validity coefficient of the set of FFQ data can be estimated from the correlation coefficients between the observed pairs of measures, as in (\ref{eqn:vc}). The fundamental reliance of the MoT on the assumption of error independence, in order for the simplification of $\Omega$ to occur, is clear.

\subsection{The method of triads: simulation study}
\label{sec:MoTsim}

A simulation study explores the performance of the MoT across a range of settings. The statistical framework from the previous section is used to generate data from sub-studies with $n = 1000$. The parameter settings are again extracted from Prentice et al. \cite{prentice2011} where the true (log) intakes are $X_i \sim N(\mu_X = \log 2141, \sigma_X^2 =(\log 14 )^2)$ and the mean vector $\mu = (0, 0, \log 1485)$. Two settings for $\Lambda$ are considered: $\Lambda = (1, 0.95, 0.9)$ reflecting the case where both the reference and FFQ have strong positive relationships with the true intake, and the case $\Lambda = (1, 0.8, 0.5)$ where the both relationships are weaker, particularly the FFQ. The diagonal of the errors' covariance matrix (\ref{eqn:SigmaMoT}) is set as $(\sigma_M^2, \sigma_R^2, \sigma_W^2) = (\alpha^2 \sigma_X^2, (\log 21)^2, (\log 28)^2)$  where the `best case scenario' of $\alpha = 0.5$ is employed.

The performance of the MoT under varying levels of measurement error correlation is considered through four settings of $\rho = (\rho_{MR}, \rho_{MW}, \rho_{RW})$: firstly, uncorrelated errors are assumed $(\rho = (0, 0, 0))$, followed by three realistic settings of weak correlation between the errors of the reference and FFQ $(\rho = (0, 0, 0.1))$, $\rho = (0, 0, 0.3)$ and $\rho = (0, 0, 0.5)$.

For each of the eight combinations of $\Lambda$ and $\rho$, 500 data sets (consisting of ($X_i, W_i, R_i, M_i)$  for $i = 1,\ldots, n)$ are generated. For each data set, the correlation between each pair of manifest variables is calculated and combined as in (\ref{eqn:vc}) to estimate the validity coefficient of the FFQ.  Given the simulation setting, the true value of the validity coefficient can be analytically derived and the performance of the MoT thus assessed.

\begin{figure}[ht]
\begin{center}
\includegraphics[width=14cm, height=9cm]{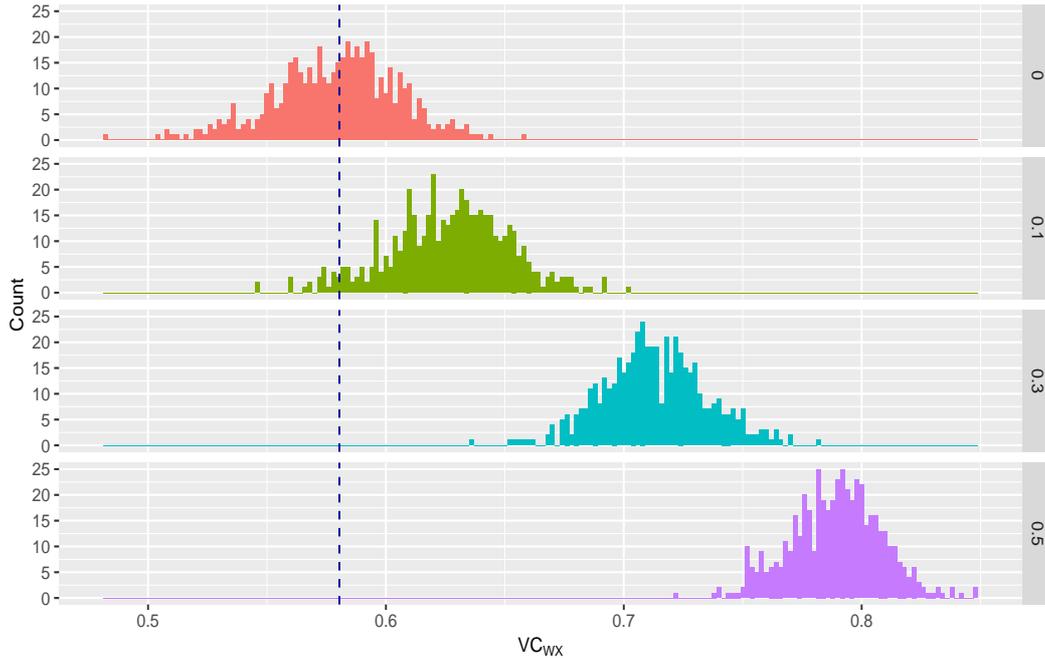}
\caption{Estimates $\hat{VC}_{WX}$ from 500 simulated data sets when $(\lambda_{RX}, \lambda_{WX}) = (0.95, 0.9)$ over the four settings for $\rho_{RW} = \{0, 0.1, 0.3, 0.5\}$, as detailed on the right hand side of each subfigure. The true value of the validity coefficient is illustrated by the dashed blue line.}
\label{fig:MoT}
\end{center}
\end{figure}

Figure \ref{fig:MoT} illustrates the estimated validity coefficient in each of 500 simulated data sets for each of the correlated error settings. For the case $(\lambda_{RX}, \lambda_{WX}) = (0.95, 0.9)$, where both the reference and FFQ are strongly and positively related to the true dietary intake, $\tilde{VC}_{WX} = 0.58$.  As expected, when the error independence assumption holds the MoT on average correctly computes the validity coefficient with mean estimate $\hat{VC}_{WX} = 0.579$. However, the MoT increasingly over-estimates the validity of the FFQ as the dependence between the errors grows: even at realistically low levels of correlation ($\rho_{RW} = 0.1$) the mean estimated validity coefficient is overestimated ($p < 0.001$). Similar observations are noted when $(\lambda_{RX}, \lambda_{WX}) = (0.8, 0.5)$ and the FFQ and reference have a weaker relationship with the true intake, as reported in Table \ref{tab:MoTest}. Also of note is the increased variability in the estimates $\hat{VC}_{WX}$ in the case of the manifest data having weaker relationship with the true intake.

\begin{table}[ht]
\caption{Mean estimates (standard errors in parentheses) of the validity coefficient across 500 simulated data sets, over four settings of $\rho_{RW}$ and across two settings of ($\lambda_{RX}, \lambda_{WX}$). The true value $\tilde{VC}_{WX}$ is detailed in the final row.}
\label{tab:MoTest}
\begin{center}
\begin{tabular}{|ll|cc|cc|}\hline
 & & & \multicolumn{2}{c}{$(\lambda_{RX}, \lambda_{WX})$} &   \\
 & & \multicolumn{2}{c}{$(0.95, 0.9)$} & \multicolumn{2}{c|}{$(0.8, 0.5)$} \\\hline
 & 0 & 0.579 & (0.026) & 0.366 & (0.034) \\
$\rho_{RW}$ & 0.1 & 0.626 & (0.025) & 0.428 & (0.033)\\
 & 0.3 & 0.713& (0.022)& 0.531 & (0.030)\\
& 0.5 & 0.789 & (0.019) & 0.617 & (0.028)\\\hline
\multicolumn{2}{|l|}{$\mathbf{\widetilde{VC}_{WX}}$} &  \multicolumn{2}{l}{$\mathbf{0.580}$} & \multicolumn{2}{|l|}{$\mathbf{0.368}$}\\\hline
 \end{tabular}
\end{center}
\end{table}

Violation of the independent errors assumption, even at low levels, notably inflates the resulting validity coefficient estimates; in such cases researchers will be over confident in the validity of their set of self-reported data. Even if good statistical practise is followed and the uncertainty in an estimate is also assessed (for example through the use of the bootstrap), the validity of  FFQ data whose errors are correlated with the utilised reference will be overestimated. In the case of suspect correlated errors, viewing the estimated validity coefficient as the upper limit of a range of possible values for the truth has been suggested \cite{dixon2006, mcnaughton2005}. The simulation studies here strongly indicate that such a practise could provide researchers with a large degree of overconfidence in their self-reported data. 

\section{Discussion and recommendations}
\label{sec:discussion}

This article set out to review and expose the state of the art methods used when combining dietary biomarkers with self-reported intake data in order to account for measurement error in dietary assessment, and to provide an exposition of the statistical detail underpinning such methods. A review of the literature highlighted two methods that are predominantly employed across a range of studies: the calibration method and the method of triads. 

The calibration method in particular is often the fundamental first step in an array of nutritional epidemiology studies, yet the statistical rigour on which the method depends is often hidden and the impact of violation of the underpinning assumptions is relatively unacknowledged. Much focus is given to the use of the results of the calibration method in subsequent models (e.g. to assess diet-disease relations in regression calibration), but the initial calibration step which combines biomarker and self-reported data often attracts little and arguably disproportionate attention. Here an overview of literature which avails of the calibration method is detailed and a synopsis of the method's statistical concepts presented. Delineating the underlying statistical model clearly exposes the centrality of the method's assumptions and a thorough simulation study explores the performance of the method when the assumptions are met or violated. Unsurprisingly, the simulation study suggests that the calibration method performs well when the method assumptions are met, particularly if the focus of the exercise is to estimate the regression coefficient relating the predicted and self-reported intakes. In such a case however, even small violations of the assumption of independence between the biomarker and self-reported data errors can lead to notable discrepancies in the coefficient estimates. On the positive side, estimation of the regression coefficient appears relatively robust to the level of variation of the biomarker around the true intake, and to a lesser extent to the strength of the relationship between the self-reported data and true intake. Somewhat conversely, if the focus of the exercise is to predict intake for study participants, predictions can in general be of similar quality in the presence of independent or correlated errors, provided the biomarker has low variance around the true dietary intake. Thus, given the challenge involved in practically establishing the presence of correlated errors, when interest lies in predicting intake focussing efforts on the development or use of a biomarker that does not vary much more than the truth, and self-reported data that are strongly related to the truth, is recommended.

The method of triads is also a well utilised approach in the nutritional science literature, in particular for establishing the validity of a set of self-reported data. The MoT provides researchers with a tool to quantify the validity of their self-reported data in terms of its correlation with the latent dietary intake. The method requires that three measurements are taken on each sub-study participant and from the empirical pairwise correlation coefficients the validity coefficient of the self-reported data can be derived. The statistical rigour underlying the method of triads is delineated herein to highlight the importance of the method's key assumption of independent errors in the three measurements. The subsequent simulation study highlights that when correlated errors do exist, even at very low levels, the validity coefficient can be significantly overestimated, providing researchers with invalid confidence in their self-reported data and in any future inference which it informs. In terms of recommendations, the results herein suggest that ensuring independence of measurement errors should be prioritised, given the influence of the violation of this assumption in comparison to the method's performance when the FFQ and reference data are only weakly representative of the truth.

While the simulation study is thorough in so far as reporting practicalities allow, it has limitations. Assumptions are made on parameter settings, and although these are made in as principled a manner as possible by deference to the literature, the parameter settings are based on an energy intake study and thus could be altered to settings more typical of, for example, a protein intake study. It would be of interest to determine if the resulting inferences and recommendations vary on a nutrient basis. Further, future simulation studies could explore the performance of the calibration method in the presence of non-classical measurement errors, in the presence of covariates or in longitudinal studies, all of which are very likely to greatly impact the method accuracy. Here, the most simple experimental setting was considered for clarity. Any concerns raised herein are likely to be exacerbated in more complex settings, such as in the presence of systematic measurement errors. 

While there is literature that concerns combining biomarker and self-reported data in the presence of correlated measurement error \cite{spiegelman2005, day2001, kaaks1994, kipnis1999, kipnis2001}, there is scope to add to this literature. The methods outlined here assume that the self-reported data are (or can be transformed to be) normally distributed; modelling the data using a heavier tailed distribution such as the t or skew-normal distribution may achieve improved model fit, and thus predicted intakes. Further, modelling self-reported data as homogeneous may also induce poor predictions in real data settings if the study population is heterogenous; considering mixtures of the latent variable models delineated herein \cite{nyamundanda2010, mcparland2014, mcparland2017}, or the related mixtures of experts models \cite{gormley2008}, may lead to strong gains when aiming to account for measurement errors in dietary assessment through combining biomarker and self-reported data. 

\section*{Acknowledgements}
This work was supported by a European Research Council grant (ERC (647783)) to LB and YB and by a Science Foundation Ireland grant (SFI/12/RC/2289) to ICG.

\bibliographystyle{Chicago}
\bibliography{GormleyBaiBrennan}

\newpage
\section*{Appendix}
\setcounter{table}{0}
\renewcommand{\thetable}{A\arabic{table}}
\vspace{-0.5cm}

\begin{longtable}{| p{1.6cm} | p{4cm} | p{4cm} | p{4cm} |}
\caption{Overview of key papers that have employed the calibration method for combining biomarkers and dietary data. (24HR: 24 hour recall, 4DFR: 4-day food record; FFQ: food frequency record, NPAAS: Nutrition and Physical Activity Assessment Study, DLW: doubly labelled water, UN: urinary nitrogen, WHI-DM: Women's Health Initiative Dietary Modification Trial, WHI-NBS: Women's Health Initiative Nutritional Biomarkers Study, CAREDS: Carotenoids and Age-Related Eye Disease Study, OPEN: Observing Protein and Energy Nutrition.)}\\\hline
\label{tab:Calib}
\textbf{Paper} & 	\textbf{Aims} & 	\textbf{Data} &	\textbf{Results/Conclusions}\\\hline\hline
\cite{kipnis2003}  $\:\:\:\:$&  
Use biomarkers to evaluate absolute protein intake, total energy and energy-adjusted protein intakes from self-reported data.  &   OPEN study $n = 484.$ \newline $\:$ \newline Dietary measurements:  FFQ and 24HR. \newline $\:$ \newline Biomarkers: DLW,  UN.  &  
Estimated attenuation factors using biomarkers as reference instrument. \newline $\:$ \newline  The use of FFQs in epidemiology is cautioned.      \\\hline
\cite{schatzkin2003} & 
Use biomarkers to compare the performance of FFQ and 24HR.  &
OPEN study $n = 484$. \newline $\:$ \newline Dietary measurements: FFQ and 24HR. \newline $\:$ \newline Biomarkers: DLW, UN. & 
FFQ is not recommended as a measurement for evaluation of relations between absolute intake of energy or protein and disease. \newline $\:$ \newline Use of multiple 24HR increased precision compared to a single 24HR, however under estimation may occur.\\\hline
\cite{neuhouser2008} & 
Use biomarkers to characterize measurement error distributions for FFQ-assessed energy and protein. \newline $\:$ \newline 
 Examine whether the measurement error structure was influenced by participant characteristics such as age, race/ethnicity or obesity.\newline $\:$ \newline Develop equations to calibrate FFQ nutrient consumption estimates & 
WHI-DM and WHI-NBS. \newline $\:$ \newline Dietary  measurement: FFQ. \newline $\:$ \newline Biomarkers:  DLW, UN. & 
Concluded that participant characteristics are important to include in calibration equations. \\\hline
\cite{freedman2011} &
Combining biomarker and dietary intake to estimate the association between diet and disease. &
Simulated data from CAREDS. \newline $\:$ \newline Dietary measurement: FFQ  \newline $\:$ \newline Biomarkers:  Serum lutein and zeaxanthin measurements (addition of lutein and zeaxanthin)  \newline $\:$ \newline True dietary intake: \newline lutein/zeaxanthin intake &
Inclusion of the biomarker in the regression calibration-estimated intake can increase statistical power and it provides nearly unbiased estimates of association between diet and disease. \\\hline
\cite{prentice2011} &
To evaluate and compare FFQ, 4DFR, and 24HRs for estimation of energy and protein intake using biomarkers. & 
NPAAS $n = 450$. \newline $\:$ \newline  Dietary measurements: FFQ, 4DFR and 24HRs. \newline $\:$ \newline  Biomarkers:  DLW, UN. & 
Developed calibration equations for the 3 self-reported methods.\newline $\:$ \newline 
Concluded that calibration equations using any of the 3 self-reported methods may yield suitable estimates.  \\\hline
\cite{tasevska2011} &
Develop a measurement error model for sugar biomarkers. \newline $\:$ \newline  
Use the biomarkers to estimate the attenuation related to intake of absolute total sugars. &
OPEN study  $n = 484$ \newline $\:$ \newline  Dietary measurements:    two FFQ and two 24HR	\newline $\:$ \newline Biomarkers:  urinary sugars (fructose + sucrose) & 
Developed a model for use of urinary sugar markers for estimated sugar intake. \\\hline
\cite{huang2013} &
Used biomarkers to correct self-reported dietary data. &
NPAAS $n = 450$. \newline $\:$ \newline  Dietary measurements:   FFQ, 4DFR and 24HR. \newline $\:$ \newline  
Biomarker:  24-hour urinary excretion assessments. & 
Simple linear calibration equations using estimates from 4DFR or three 24HRs can capture much of the variation in usual daily nutrient consumption. 
\newline $\:$ \newline
The FFQ based dietary data had limited ability in this regard. \\\hline
\cite{mossavar2013} & 
Examine the impact of psychosocial and diet behaviour on measurements errors of self-reported data. & 
NPAAS postmenopausal women, $n = 450$.  \newline $\:$ \newline Dietary measurements: FFQ, 4DFR and 24HRs.  \newline $\:$ \newline Biomarkers:  DLW, UN.&
The contribution of the examined parameters was modest in comparison to that of BMI, age and ethnicity. \\\hline
\cite{prentice2013} & 
Assessing the relationship between fat and total energy intake with postmenopausal breast cancer.&
NPAAS,  $n = 450$.\newline $\:$ \newline Dietary measurements: FFQ, 4DFR and 24HRs. \newline $\:$ \newline Biomarkers: 
DLW. &	
Calibrated total energy intake was positively associated with postmenopausal breast cancer incidence. \newline $\:$ \newline 
The association was not evident without biomarker calibration. \\\hline
\cite{freedman2014} &
Examination of reporting errors in FFQ and 24HR and characteristics associated with such errors. &	
Pooled data from 5 large validations studies. \newline $\:$ \newline Dietary measurements: FFQ and 24HR. \newline $\:$ \newline Biomarkers: DLW, UN.	&
Calibration equations for true intake that included personal characteristics improved prediction. \\\hline
\cite{tasevska2014} & 
Assess estimation of total sugar intake from 3 self-reported tools against 24h urinary sugars. \newline $\:$ \newline 
Development of calibration equations that predict total sugars intake. &	
NPAAS $n = 450$. \newline $\:$ \newline Dietary measurements:  FFQ, 4DFR and 24HR. \newline $\:$ \newline 
Biomarkers: urinary sugars (fructose + sucrose).&
None of the self-reported instruments provided a good estimate of sugars intake. \newline $\:$ \newline 
Measurement of biomarkers in sub samples may be necessary for calibration of the data. \\\hline
\cite{freedman2014} &
Examination of reporting errors in FFQ and 24HR and characteristics associated with such errors. &	
Pooled data from 5 large validations studies.  \newline $\:$ \newline Dietary measurements: FFQ and 24HR.  \newline $\:$ \newline Biomarkers: DLW, UN.&	
Calibration equations for true intake that included personal characteristics improved prediction. \\\hline
\end{longtable}

\newpage
\begin{longtable}{| p{1.89cm} | p{4cm} | p{4cm} | p{4cm} |}
\caption{Selection of studies using the method of triads with biomarker data as a reference method. (24HR: 24 hour recall, 7DFR: 7-day food record; FFQ: food frequency record, EPIC: European Prospective Investigation into Cancer and Nutrition study, EFCOVAL: the Dutch participants of the European Food Consumption Validation.)}\\\hline
\label{tab:MoT}
\textbf{Paper} & 	\textbf{Aims} & 	\textbf{Data} &	\textbf{Results/Conclusions}\\\hline\hline
\cite{kaaks1997} &
Compare and evaluate replicate measurements and including biomarker measurements. 	&
EPIC $n = 521 000$.\newline $\:$ \newline Dietary measurements:  FFQ and 24HR. \newline $\:$ \newline 
Biomarkers: Serum Vitamin C.  &
Use method of triads with FFQ, 24HR and biomarker to estimate the validity coefficient. \newline $\:$ \newline 
Using biomarkers in dietary validity studies can make it more likely that the criteria of independent errors, crucial in validity studies are met.\\\hline
\cite{kabagambe2001}&
Use of method of triads to validate FFQ and biomarkers. 	&
Validation study from study of myocardial infarction in Costa Rica $n = 120$.\newline $\:$ \newline Dietary measurements: 
two FFQs and seven 24HR. \newline $\:$ \newline Biomarkers: 
Plasma tocopherol and carotenoids.  Adipose tissue tocopherol, carotenoid and fatty acids.&	
Used the method of triads to validate the use of FFQ in a Hispanic population. \\\hline
\cite{rosner2008} &
Estimate the regression coefficients between true intake and self-reported when the measurement error in self-reported and reference are correlated using the method of triads.	&
Vitamin C data from EPIC-Norfolk study $n = 322$.\newline $\:$ \newline Dietary measurements:  FFQ and 7DFR \newline $\:$ \newline Biomarkers: Plasma vitamin C. &
Presented an extension that allows for the presence of correlated error between a surrogate instrument and a gold standard in a longitudinal setting. \\\hline
\cite{geelen2015} &
Illustrate the impact of intake-related bias in FFQ and 24HR, and correlated errors between these self-reports, on intake-health associations. & 
EFCOVAL $n = 122$.  \newline $\:$ \newline Dietary measurements:  24HR and FFQ. \newline $\:$ \newline Biomarkers: 
Urinary sodium and potassium. &
De-attenuation using the method of triads and other methods with duplicate recovery biomarkers and duplicate 24HR. \newline $\:$ \newline Calibration of the FFQ intake data to a gold standard reference method is preferred. \newline $\:$ \newline 
Correlated errors between FFQ and 24HR limit the use of the validity coefficient as a correction factor. \\\hline
\end{longtable}

\end{document}